\documentclass[11pt]{article}
\baselineskip
\pdfoutput=1
\usepackage{amsmath}
\usepackage{amssymb}
\usepackage{graphicx}
\usepackage{color}
\usepackage{cite}
\usepackage{collref}
\usepackage{tabls}
\usepackage{hyperref}
\usepackage{diagbox}
\usepackage{float} 
\usepackage{cancel}
\DeclareMathOperator{\Tr}{Tr}
\newcommand{\dd}{{\rm{d}}}
\newcommand{\DD}{{\rm{D}}}
\newcommand{\tmb}[1]{{\mbox{\tiny{#1}}}}

\newcommand{\eq}{\begin{equation}} 
\newcommand{\en}{\end{equation}} 
\newcommand{\eqa}{\begin{eqnarray}}
\newcommand{\ena}{\end{eqnarray}}

\begin {document}

\begin{titlepage}

\begin{center}
{\Large\bf Sampling the lattice Nambu-Goto string using \\ Continuous Normalizing Flows}
\end{center}
\vskip1.3cm
\centerline{Michele~Caselle,\footnote{\href{mailto:caselle@to.infn.it}{{\tt caselle@to.infn.it}}} Elia~Cellini,\footnote{\href{mailto:elia.cellini@unito.it}{{\tt elia.cellini@unito.it}}} and Alessandro~Nada\footnote{\href{mailto:alessandro.nada@unito.it}{{\tt alessandro.nada@unito.it}}}}
\vskip1.5cm
\centerline{\sl Department of Physics, University of Turin and INFN, Turin}
\centerline{\sl Via Pietro Giuria 1, I-10125 Turin, Italy}
\vskip1.0cm

\setcounter{footnote}{0}

\begin{abstract}
\noindent
Effective String Theory (EST) represents a powerful non-perturbative approach to describe confinement in Yang-Mills theory that models the confining flux tube as a thin vibrating string.
EST calculations are usually performed using the zeta-function regularization: however there are situations (for instance the study of the shape of the flux tube or of the higher order corrections beyond the Nambu-Goto EST) which involve observables that are too complex to be addressed in this way. In this paper we propose a numerical approach based on recent advances in machine learning methods to circumvent this problem. Using as a laboratory the Nambu-Goto string, we show that by using a new class of deep generative models called Continuous Normalizing Flows it is possible to obtain reliable numerical estimates of EST predictions.
\end{abstract}

\end{titlepage}

\section{Introduction}

In the last few years Effective String Theory (EST) has emerged as a highly promising approach for the understanding and modeling of the non-perturbative behavior of confining Yang-Mills theories. In this framework, the confining flux tube that connects a quark-antiquark pair is represented as a thin vibrating string~\cite{Nambu:1974zg, Goto:1971ce, Luscher:1980ac, Luscher:1980fr, Polchinski:1991ax}.

In $D\neq 26$, where $D$ are the space-time dimensions of the target Lattice Gauge Theory (LGT), the EST (at least in its simplest formulation) is anomalous at the quantum level and thus must be considered only as an effective, large-distance description of Yang-Mills theories. Notwithstanding this, precise Monte Carlo simulations of several different LGTs proved that it is indeed a highly predictive effective model (for recent reviews see for instance~\cite{Aharony:2013ipa, Brandt:2016xsp, Caselle:2021eir}).

The reason of this success is in the so called "low energy universality"~\cite{Meyer:2006qx, Luscher:2004ib, Aharony:2009gg, Aharony:2011gb, Dubovsky:2012sh, Billo:2012da, Gliozzi:2012cx, Aharony:2013ipa} which states that, due to the symmetry constraints imposed by the Poincar\'e invariance in the target space, the first few terms of the EST large-distance expansion are universal and coincide with those of the Nambu-Goto action~\cite{Nambu:1974zg, Goto:1971ce}.  As a consequence, the EST turns out to be much more predictive than typical effective models and its predictions depend only on one free parameter: the string tension $\sigma$ of the Nambu-Goto model.

It is thus clear that a central role in this game is played by the Nambu-Goto action. Its predictions can be immediately compared with the results of Monte~Carlo simulations of essentially any confining LGT (we shall comment below on a few relevant exceptions to this statement). Thus, a precise understanding of its physical properties may lead to a better understanding of the confining regime of Yang-Mills theories.   

A major progress in this context was the recent observation that the Nambu-Goto model is actually an exactly integrable, irrelevant, perturbation of the two-dimensional free Conformal Field Theory (CFT)~\cite{Dubovsky:2012sh} of $D-2$ free bosons which represent the transverse degrees of freedom of the string. This irrelevant perturbation is driven by the $T\bar T$ operator of the $D-2$ free bosons~\cite{Caselle:2013dra, Smirnov:2016lqw, Cavaglia:2016oda}.

We shall see below that several results are analytically known for the partition function of the Nambu-Goto action, which can be explicitly calculated for essentially all the geometries which are relevant for the comparison with LGT results. On the contrary, much less is known on the correlation functions, and in particular on the one that measures the density of the chromoelectric flux in presence of quark sources: this quantity has an important meaning in LGT, since it allows to study the shape of the flux tube.

In principle one could address this problem, by treating the Nambu-Goto action as a spin model and regularizing it on a two-dimensional lattice, where one could perform Monte~Carlo simulations. However, due to the strong non-linearity of the model, it turns out that performing such simulations using standard algorithms is highly inefficient. The problem is instead perfectly suited for a completely different numerical approach, based on deep learning architectures.

In this respect it is important to stress that the spin model that we obtain discretizing the Nambu-Goto action is critical for the whole set of values of the string tension. Its continuum limit is in fact, as mentioned above, the theory of a free bosonic massless degree of freedom perturbed by the $T\bar T$ operator. This makes this model a perfect laboratory to test different algorithms in a particularly challenging setting.

In recent years, the exponential growth of machine learning has stimulated the introduction of novel methods for numerical studies of quantum field theory~\cite{Radovic:2018dip, Carleo:2019ptp, Carrasquilla:2020mas, Boehnlein:2021eym, Schwartz:2021ftp, Dawid:2022fga, Boyda:2022nmh, Zhou:2023pti}. A very promising class of algorithms are Normalizing Flows (NF)~\cite{rezende2015variational}, deep generative models used to efficiently sample from complex statistical distributions. NFs provide highly flexible, invertible and differentiable maps between suitable base distributions and the target. One possible application of such models is to sample configurations from Boltzmann distributions: this approach found successful application in toy models of lattice field theory~\cite{Albergo1, Nicoli2021, Albergo2, Albergo3, Albergo4, Hackett:2021idh, Abbott:2022zhs, DelDebbio, Pawlowski:2022rdn, Lawrence:2021izu, Caselle:2022acb,Lawrence:2022afv, Finkenrath:2022ogg, deHaan:2021erb, Gerdes:2022eve, Chen:2023ff, Bacchio:2022, Albandea:2023wgd, Nicoli:2023qsl, Abbott:2023thq,Singha:2023xxq}. While standard NFs still suffer from poor scaling~\cite{Abbott:2022zsh}, a generalization of these algorithms called Continuous Normalizing Flows (CNFs)~\cite{chen:2018} has been used to obtain interesting results in lattice scalar field theory~\cite{deHaan:2021erb, Gerdes:2022eve, vaitl:2022}. 

In this paper we used a CNF architecture to evaluate the partition function and the shape of the flux tube of a lattice-regularized version of the Nambu-Goto model. Thanks to the exact knowledge of the partition function we can benchmark the correctness of this novel numerical approach and then use the results to obtain information on the shape of the flux tube. This rather unusual approach to EST calculations is interesting for several reasons that we discuss in the following.

\begin{itemize}
\item It allows to study large values of $\sigma$ which are not accessible in LGT simulations; they are of great interest since they correspond to the perturbative regime of the $T\bar T$ perturbation which is controlled by $1/\sigma$~\cite{Smirnov:2016lqw, Cavaglia:2016oda}.
\item  On the LGT side, it allows for a precise study of the flux-tube shape predicted by the Nambu-Goto model, which can be used as a benchmark to compare results from LGT simulations. In this context, for instance, we will confirm the predictions of~\cite{Gliozzi:2010jh, Gliozzi:2010zt, Gliozzi:2010zv}.
\item It is the first step toward an analytic study of the corrections beyond Nambu-Goto, which have been recently the subject of great interest in the EST community~\cite{EliasMiro:2019kyf, Caristo:2021tbk, Athenodorou:2011rx, Dubovsky:2014fma, Chen:2018keo, Baffigo:2023rin}.
\item With a suitable modification of the Lagrangian of our model it would be possible to study, with this same methods, the so-called "rigid string", for which very few analytic results are known. This would provide the essential missing ingredient to test if the rigid string is the correct model to describe theories, such as the three-dimensional $U(1)$ model~\cite{Caselle:2014eka}, in which confinement is due to monopole condensation~\cite{Polyakov:1976fu}.  
\item It allows to study in detail the role of lattice artifacts in the Nambu-Goto predictions for LGTs.
\item Last, but not least, this is the first numerical study of the lattice realization of a $T\bar T$ perturbed model and could provide a tool to test in a non-trivial setting various predictions obtained in the last few years on this class of models.
\end{itemize}

While some of these goals are for the moment beyond the scope of this contribution, this work represents a proof of concept for the feasibility of this approach. As a test of the efficiency of this approach we numerically evaluate the next-to-leading order of the flux tube width, for which no Monte Carlo result in LGT had been obtained up to now.

This paper is organized as follows. In section~\ref{sec:NG} we will briefly recall a few basic results on the Nambu-Goto model, then in section~\ref{sec:CNF} we will describe the technical aspects of Continuous Normalizing Flows and how we obtained our numerical results, which we describe in detail in section~\ref{sec:results}. The last section will be devoted to some concluding remarks. 


\section{Nambu-Goto string}
\label{sec:NG}

We shall recall here only a few basic results on EST and in particular on the Nambu-Goto string. We refer the interested reader to the reviews~\cite{Aharony:2013ipa, Brandt:2016xsp, Caselle:2021eir} for a more complete discussion.

In EST, the quark-antiquark potential is modelled in terms of a vibrating string. In particular, the correlator between two Polyakov loops is related to the sum over all the surfaces bordered by the two Polyakov loops weighted by the EST action:
$$\langle P(0) P^{\dagger}(R)\rangle \sim \int \DD X \; e^{-S_\tmb{EST}[X]}$$
where $R$ denotes the distance between the two Polyakov loops.

The simplest choice for $S_\tmb{EST}$ fulfilling the constraints imposed by the Lorentz invariance in the target space is 
the Nambu-Goto action~\cite{Nambu:1974zg,Goto:1971ce} that is defined as follows:
\begin{align}
\label{NGaction}
S_\tmb{NG} = \sigma \int_\Sigma d^2\xi \sqrt{g},
\end{align} 
where $g\equiv \det g_{\alpha\beta}$ and
\begin{align}
\label{NGaction2}
g_{\alpha\beta}=\partial_\alpha X_\mu~\partial_\beta X^\mu
\end{align} 
is the  metric induced on the reference world-sheet surface $\Sigma$ by the mapping $X_\mu(\xi)$ of the worldsheet in the target space,
and $\xi\equiv(\xi^0,\xi^1)$ denote the worldsheet coordinates. This term has a simple geometric interpretation: it measures the area of the surface spanned by the string in the target space and has only one free parameter: the string tension $\sigma$.

$S_\tmb{NG}$ is manifestly reparametrization-invariant and the first step is to fix this invariance. The standard choice is the so-called ``physical gauge''. In this gauge the two worldsheet coordinates are identified with the longitudinal degrees of freedom of the string: $\xi^0=X^0$, $\xi^1=X^1$, so that the string action can be expressed as a function only of the $(D-2)$ degrees of freedom corresponding to the transverse displacements, $X^i$, with $i=2, \dots , (D-1)$ which are assumed to be single-valued functions of the worldsheet coordinates. In the following, for simplicity, we assume $D=3$ so as to have only one transverse direction. We thus eliminate the index $i$ and simply denote as $X(\xi)$ the remaining degrees of freedom. 

It is well known that this gauge fixing is anomalous at the quantum level and the resulting action should only be considered as a large distance approximation of the true string action. This is the reason why we define this approach as an "effective" string description of confinement. With a suitable redefinition of the fields, $X=\phi/\sqrt{\sigma}$, we can write explicitly this large distance expansion as follows 
\begin{equation}
S_\tmb{NG}=\sigma R L+\frac{1}{2}\int d^2\xi\left[\partial_\alpha \phi\cdot\partial^\alpha \phi - \frac{1}{8\sigma R L}(\partial_\alpha \phi \cdot\partial^\alpha \phi)^2 +\cdots \right].
\label{action2NG}
\end{equation}
where $R$ denotes, as above, the distance between the two Polyakov loops and $L$ their length.
The first term of this expansion is exactly the gaussian action, i.e. a two dimensional Conformal Field Theory (CFT) of a free bosonic field which represents the only remaining transverse degree of freedom of the string in $D=3$. Remarkably enough, all the remaining terms of the expansion combine themselves to give an exactly integrable, irrelevant perturbation of this CFT~\cite{Dubovsky:2012sh}, driven by the $T\bar T$ operator of the $D-2$ free bosons~\cite{Caselle:2013dra, Smirnov:2016lqw, Cavaglia:2016oda}.
Using the zeta function regularization it is possible to evaluate exactly the partition function of this CFT to all orders in this irrelevant perturbation. The result is~\cite{Luscher:2004ib, Billo:2005iv}

\begin{equation}
\langle P(x)P^\dagger(x+R) \rangle 
  \sim \sqrt{\sigma}{L}
   \sum_{n=0}^{\infty}w_n
  K_{0}({E}_nR)
\label{NG}
\end{equation}
where $K_{0}$ is the modified Bessel function of order $0$, $w_n$ is the multiplicity of the closed string states which propagate from one Polyakov loop to the other, and $E_n$ their energies:
\begin{equation}
  {E}_n(L)=\sigma L
  \sqrt{1+\frac{8\pi}{\sigma L^2}\left[n-\frac{1}{24}\right]}.
\label{energylevels}
\end{equation}
From the Polyakov loop correlator it is possible to obtain the interquark potential, which is defined as follows
\begin{equation}
 V(R,L)=-\frac{1}{L}\log{\langle P(x)P^\dagger(x+R) \rangle }
 \end{equation}
and in the large-$R$ limit we have, as expected, a linearly rising potential whose slope is controlled by the ground state energy of the EST
\begin{equation}
 V(R,L) = \frac{E_0(L)}{L} R = \sigma \sqrt{1-\frac{\pi}{3\sigma L^2}} R
\end{equation}

In the following for our comparisons we will perform an expansion in powers of $\frac{1}{\sigma R L}$ of the action. The first few terms of this expansion were actually obtained by Dietz and Filk~\cite{Dietz:1982uc} much earlier than eq.~(\ref{NG}), by treating the next-to-leading terms which appear in eq.~(\ref{action2NG}) as small perturbations of the gaussian free action.
In particular the first term is the well-known partition function of the free bosonic $c=1$ CFT
\begin{equation}
\label{loop1}
\langle P(x)P^\dagger(x+R) \rangle \sim \frac{e^{-\sigma R L }}{ \eta(\tau)}
\end{equation}
where $\eta(\tau)$ denotes the Dedekind $\eta$ function 
\begin{equation}
\eta(\tau)=q^\frac{1}{24}\prod_{n=1}^\infty(1-q^n)\hskip0.5cm
;\hskip0.5cmq=e^{2\pi i\tau}
\hskip0.5cm
;\hskip0.5cm \tau=i\frac{L}{2R}~~~.
\label{eta}
\end{equation}
\vskip.3cm

To understand the meaning of this result it is useful to expand it in the two limits $R\ll L$ and $R\gg L$, in which we can neglect the $q^n$ terms in the infinite product of the Dedekind function. These limits correspond in the LGT language to the low- and high-temperature limits of the confining regime\footnote{To avoid confusion let us stress that with "high-temperature" we mean a region near the deconfinement transition, but still in the confining phase.} while from a string point of view they correspond to the open and closed string channels respectively:
\begin{equation}
V(R,L)=\sigma R -\frac{\pi }{24 R} \qquad \qquad \mbox{for } R\ll L
\label{eta1bis}
\end{equation}

\begin{equation}
V(R,L)=\sigma R -\frac{\pi R}{6 L^2}+\frac{1}{2L} \log\frac{2R}{L} \qquad \qquad \mbox{for } R\gg L
\label{eta2bis}
\end{equation}

The $-\frac{\pi }{24 R}$ term in the first limit is the well known "L\"uscher term" which has been widely studied in the LGT context. The next-to-leading term (i.e. the $1/\sigma R L$ correction) can be recast in the following combination of the Eisenstein functions $E_2$ and $E_4$~\cite{Dietz:1982uc}:
\begin{equation}
\langle P(x)P^\dagger(x+R) \rangle 
  \sim \frac{e^{-\sigma R L }}{ \eta(\tau)} \left(1+\frac{\pi^2 L}{1152\sigma R^3}
  \left[ 2E_4(\tau)-E_2^2(\tau)  \right]  \right)
\label{loop2}
\end{equation}
where $E_2$ and $E_4$ can be expressed in power series
\begin{align}
E_2(\tau) &= 1 - 24\sum_{n=1}^\infty \sigma(n) q^n\\
E_4(\tau) &= 1 + 240\sum_{n=1}^\infty \sigma_3(n) q^n
\end{align}
and $\sigma(n)$ and $\sigma_3(n)$ are, respectively, the sum of all divisors of $n$ (including 1 and $n$) and the sum of their cubes.

For the range of values of $\sigma$, $R$ and $L$ that we shall study in this paper the terms proportional to $q$ can be systematically neglected and we end up with the following expressions in the two limits:

\begin{equation}
V(R,L)=\sigma R -\frac{\pi }{24 R} - \frac{\pi^2 L}{1152\sigma R^3}+\cdots \qquad \qquad \mbox{for } R\ll L
\end{equation}

\begin{equation}
V(R,L)=\sigma R +\frac{1}{2} \log\frac{2R}{L}-\frac{\pi R}{6 L^2}-\frac{\pi^2 R}{72 L^4} +\cdots \qquad \qquad \mbox{for } R\gg L
\label{eta2}    
\end{equation}

It is important to stress the role in these calculations of the zeta function regularization, which eliminates all the "bulk", divergent terms of the above partition functions. We shall further comment on this point below.

\subsection{Width of the flux tube according to the Nambu-Goto action}
\label{NGwidth}

In the EST framework the width of the flux tube in the position $(\xi^0,\xi^1)$ is given by
\eq
  w^2(\xi;R,L)=\left\langle \phi^2(\xi^0,\xi^1)\right\rangle
  \label{w3}
\en
where $R,L$ denote the distance between the Polyakov loops and their length. This correlator is singular and must be regularized: the most natural choice is a point-splitting regularization:
\eq
  w^2(\xi;R,L)=\left\langle \phi(\xi^0,\xi^1)\phi(\xi^0+\epsilon,\xi^1+\epsilon)\right\rangle
  \label{w3bis}
\en

Due to the periodic boundary condition in the time direction the result must be translation-invariant in that direction and hence we do not expect any dependence on $\xi^0$. Moreover, following the usual convention in LGT we fix $\xi^1$ to be exactly the midpoint between the two Polyakov loops $\xi^1=R/2$, so that the EST width is a function only of $R$ and $L$.

As we mentioned in the introduction much less is known analytically on $w^2$ with respect to the partition function discussed previously. In particular, no result is known for the width using the whole Nambu-Goto action and obtaining a numerical estimate of this result is one the goals of our line of research. An explicit expression can be obtained for the free Gaussian action~\cite{Luscher:1980iy} (see also~\cite{Caselle:1995fh, Allais:2008bk} for a detailed calculation of the width in the cylindric geometry in which we are presently interested)  
\eq
\sigma w^2(z)=-\frac{1}{2\pi} \log\frac{ \pi |\epsilon|}{2 R} + 
\frac{1}{2\pi} \log\left|\theta_2\left(0\right)/\theta_1'(0) \right|
\label{ris1}
\en
where $\theta_i$ are the Jacobi $\theta$ functions defined as:
$$
\theta_1(z)=2q^{\frac14}\sum_{n=0}^{\infty}(-1)^n q^{n(n+1)} \sin(2n+1)z
$$
and
$$
\theta_2(z)=2q^{\frac14}\sum_{n=0}^{\infty} q^{n(n+1)} \cos(2n+1)z
$$
with
$$
q=e^{-\pi L/2R}.
$$
This expression simplifies in the limits $L\gg R$ and $R\gg L$ in which we are interested:
\begin{itemize}
\item {\bf $L\gg R$}

In this case, setting $R_c= \pi |\epsilon|/2$ we find the well-known logarithmic increase of the flux tube width with the interquark distance $R$
\eq
\sigma w^2(z)= \frac{1}{2\pi} \log\frac{R}{R_c}
\en
\item
{\bf $L\ll R$} 
 
In this case, setting $L_c= \pi |\epsilon|$ we have instead a linear increase of the width as a function of $R$
\eq
  \sigma w^2(z)=\frac{1}{2\pi} \log\frac{L}{L_c} +
  \frac{R}{4L}+\cdots
\label{risfinale}
\en
\end{itemize}

Remarkably enough it is also possible to obtain the next to leading contribution to $w^2$~\cite{Gliozzi:2010jh, Gliozzi:2010zt, Gliozzi:2010zv}. Also in this case the expression simplifies in the two limits and one finds:
\begin{itemize}
\item {\bf $L\gg R$}
\eq
\sigma w^2(z)= \frac{1}{2\pi} \log\frac{R}{R_c} \left(1-\frac{\pi}{4\sigma R^2}\right)+ \frac{5}{96} \frac{1}{\sigma R^2}
\label{GPLT}
\en

\item
{\bf $L\ll R$} 
\eq
  \sigma w^2(z)=\frac{1}{2\pi} \log\frac{L}{L_c} +
  \frac{R}{4L} + \frac{\pi}{24}\frac{R}{\sigma L^3}+\cdots
\label{GPHT}
\en
\end{itemize}
These are the results that we shall compare with our numerical simulations.

\subsection{Lattice regularization of the Nambu-Goto action}

As we mentioned in the introduction, our strategy is to treat the Nambu-Goto model as a two-dimensional spin model and then study it numerically. To this end the first step is to discretize the model on a two-dimensional lattice representing the worldsheet of the string, which is not to be confused with the $D-$dimensional lattice in which the original LGT lives. Following the usual convention we subtract from the Nambu-Goto action the bulk term proportional to the area and rewrite it as:
\begin{equation}\label{eq:NG}
    S_\tmb{NG}=\sigma \sum_{x \in \Lambda} \biggl(\sqrt{1+(\partial_{\mu}\phi(x))^2/\sigma}-1\biggr) 
\end{equation}
where $\Lambda$ is a square lattice of size $L\times R$ with index $x=(x^0,x^1)$ representing the worldsheet of the string\footnote{We rescaled the worldsheet coordinates in this section to reabsorb the $RL$ factors in the subleading terms of the action so as to make  the various terms of the exact solution on the lattice more explicit.} and lattice step $a=1$; $\phi(x) \in \mathbb{R}$ is a real scalar field representing the transverse degrees of freedom of the string. Along the temporal extension of length $L$ we fix periodic boundary conditions $\phi(x^0,x^1)=\phi(x^0+L,x^1)$; thus, we can identify the physical temperature as the inverse of $L$. Along the spatial extension of length $R$ we fix Dirichlet boundary conditions $\phi(x^0,0)=\phi(x^0,R)=0$, which represent the Polyakov loops $P$ and $P^{\dagger}$. The only coupling constant of the model is the string tension $\sigma$.

As in the previous section we can expand eq.~(\ref{eq:NG}) in powers of $1/\sigma$:
\begin{equation}\label{eq:Expansion}
    S_\tmb{NG} \sim S_\tmb{FB}+O(\sigma^{-1}) 
\end{equation}
where
\begin{equation}
    S_\tmb{FB}[\phi]=\frac{1}{2}\sum_{x \in \Lambda}(\partial_{\mu}\phi(x))^2
\end{equation}
is the lattice discretization of the free boson action.

The lattice discretization has two main effects: the first is the appearance of a set of "bulk" constants, which diverge in the continuum limit\footnote{These constants are proportional to $L$ (a "bulk" boundary term due to the Polyakov loops) or to $RL$ (area term). Due to the periodicity in the time direction, no term proportional to $R$ can appear.}, the second is the appearance of finite size corrections which are proportional to negative powers of the lattice sizes $R$ and $L$ and vanish in the continuum limit\footnote{Since we are keeping the lattice spacing fixed to $a=1$ the continuum limit corresponds to the $R,L\to \infty$ limit.}.
Both sets of terms are not universal, but depend on the details of the lattice regularization and are automatically eliminated by the zeta function regularization which selects only the adimensional terms in the expansion. These adimensional terms are the only ones which are "universal" in the renormalization group sense and survive in the continuum limit. Our main goal in the following will be the comparison of our numerical results with the zeta function predictions for these universal terms.
  
While the non-universal terms are in principle undetermined, with a few simple choices for the lattice regularization, the contribution to the partition function of $S_\tmb{FB}$ can be evaluated exactly also on a finite lattice, thus allowing to use them too as benchmarks of our calculations.
  
We sketch here the main steps of the solution (see also~\cite{Caselle:1996kd} for further details).
The lattice action can be diagonalized with eigenfunctions
\begin{equation}\label{eq:equigenfunction}
    \Psi(x^0, x^1,m,n)=\frac{2}{\sqrt{2LR}} \biggl( \cos\biggl(\frac{2m\pi x^0}{L}\biggr)+\sin\biggl(\frac{2m\pi x^0}{L}\biggr) \biggr)\sin\biggl(\frac{n\pi x^1}{R}\biggr)
\end{equation}
and eigenvalues
\begin{equation}\label{eq:eigenvalues}
    \lambda_k\equiv\lambda_{m,n}= 4 \sin^{2}\biggl(\frac{m\pi}{L}\biggr) + 4\sin^{2}\biggl(\frac{n\pi}{2R}\biggr).
\end{equation}

Performing the Gaussian integrations we obtain 
\begin{equation}
    Z_\tmb{FB}=\int D\phi e^{-S_\tmb{FB}[\phi]}=\prod_{m=1,n=1}^{L,R-1} \sqrt{ \frac{2\pi}{\lambda_{m,n} }}
\end{equation}
or equivalently
\begin{equation}
\label{zfb}
  -\log(Z_\tmb{FB})= -\frac12 \sum_{m=1,n=1}^{L,R-1} \log{\left({ \frac{2\pi}{\lambda_{m,n} }}\right)}.
\end{equation}
 
This sum is divergent and is usually regularized with the zeta function. However, as mentioned above, in the present case we are interested also in these non-universal terms, because they will appear in the simulation of the whole Nambu-Goto action, and we can use the exact solution of the free bosonic part to fix them.

Indeed the sum in eq.~(\ref{zfb}) can be evaluated numerically with any precision and besides the Dedekind function one finds two divergent contributions, which are proportional respectively to the area $RL$ and to the length of the boundary $L$:
\begin{equation}
  -\log Z_\tmb{FB}= A_\tmb{FB}~ RL + C_\tmb{FB}~L + \log{\eta(\tau)}
  \label{zfb1}  
\end{equation}
with
\begin{equation}
  A_\tmb{FB}=-0.3358177..,\hskip 1cm C_\tmb{FB}=0.478252... 
  \label{zfb2}
\end{equation}
being the constants which we shall use in section~\ref{sec:results} to benchmark our numerical results.

\begin{figure}[H]
  \centering
\includegraphics[scale=1.0,keepaspectratio=true]{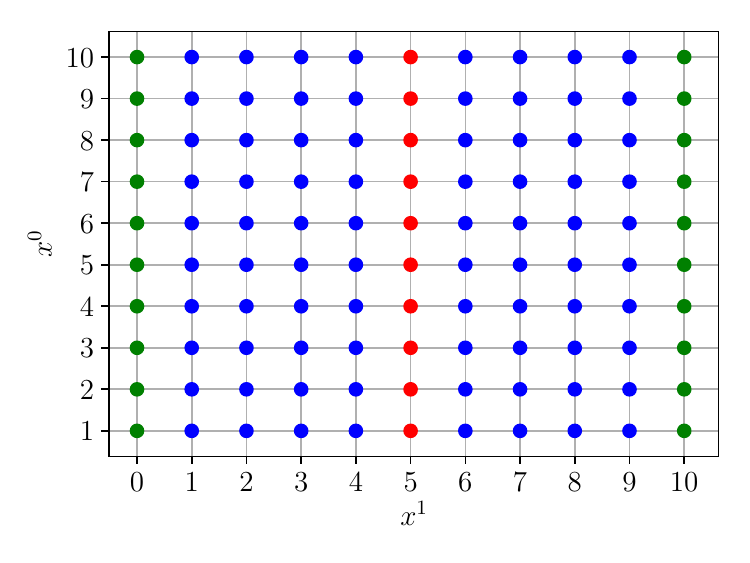}
  \caption{Schematic representation of the lattice configurations $\phi$ with volume $L\times R=10\times 10$. The green sites represent the Dirichlet boundary and they are fixed to 0, blue and red sites represent the "active" volume seen by the CNFs. The width $\sigma w^2$ of one configuration $\langle \phi^2(x^0,R/2)\rangle_{x^0}$ is computed by averaging over $x^0$ the square of the red sites.}
   \label{fig:ltcConf}
 \end{figure}

The width of the regularized Nambu-Goto string can be expressed as:
\begin{equation}
 \sigma w^2(\sigma, L,R)=\langle \phi^2(x^0,R/2)\rangle_{x^0}    
\end{equation}
where the expectation value $\langle ... \rangle_{x^0}$ is computed also by averaging over the temporal dimension $x^0$ (due to the translation invariance imposed by the periodic boundary conditions). In fig.~\ref{fig:ltcConf}, a schematic representation of the configurations $\phi$ is reported.

\subsection{Implications of our approach for EST modelling.}
The increase in precision of simulations in Lattice Gauge Theories is posing new challenges to EST models. In particular it is by now clear that the Nambu-Goto action should be considered only as a first approximation of the actual EST describing the confining regime of LGTs\cite{EliasMiro:2019kyf, Caristo:2021tbk, Athenodorou:2011rx, Dubovsky:2014fma, Chen:2018keo, Baffigo:2023rin}. In order to address more complex models one should first have a perfect control of the Nambu-Goto EST.  This has been achieved in the past years for the partition function of the model but it is far to be reached for the correlation functions and in particular for the width of the flux tube. This width, and more generally the shape of the flux tube is one of the main observables of interest in LGT simulations and has been shown to discriminate among different confining mechanisms~\cite{Caselle:2014eka}. For instance, we expect a Gaussian-like shape for the Nambu-Goto action while in presence of a term (beyond the Nambu-Goto action) proportional to the extrinsic curvature of the string  we expect an exponentially decreasing shape, with the appearance of a new scale similar to the London penetration length of the Abriksov vortices. The main analytical tool we have to address these problems is the zeta function regularization, but, as we mentioned above, there are situations in which it cannot be used due to the complexity of the observable or to the non-perturbative nature of the corrections.  Our approach can be used to overcome these problems and in fact, as we shall see below, we obtain for the first time a precise determination of the next to leading term of the flux tube width for the Nambu-Goto EST. This is a preliminary step toward a similar analysis in more complex models and, possibly, will give us a tool to discriminate among different confining mechanisms.
 
\section{Continuous Normalizing Flows}
\label{sec:CNF}

Normalizing Flows~\cite{rezende2015variational,papamakarios2021normalizing} are a deep learning architecture based on invertible and differentiable transformations $f: \mathbb{R}^V\to\mathbb{R}^V$ which interpolate between a prior distribution $q_0(z), \; z \in \mathbb{R}^V$ and a target distribution, which in our case is set to be the Boltzmann distribution $p(\phi)=\exp(-S_\tmb{NG}[\phi])/Z,\;\phi \in \mathbb{R}^V$. 

Continuous Normalizing Flows represent a particular implementation of such architectures: they can be constructed setting $f$ to be the solution of a Neural Ordinary Differential Equation (NODE)~\cite{chen:2018,deHaan:2021erb, Gerdes:2022eve,vaitl:2022} in time $t\in [0,T]$
\begin{equation}
    \frac{\dd \phi(t)_x}{\dd t}=g_\theta(\phi(t),t)_x \quad \textrm{with} \quad z\equiv\phi(0) \quad \textrm{and} \quad \phi\equiv\phi(T)
\end{equation}
where $x$ is a site of the lattice $\Lambda$. The probability density $q$ of the generated samples $\phi$ can be readily computed solving the ODE
\begin{equation}
    \frac{\dd \log q(\phi(t))}{\dd t}=-\bigl(\nabla \cdot g_\theta)(\phi(t),t).
\end{equation}

In this work, we consider an architecture inspired by~\cite{Gerdes:2022eve}. The vector field $g_\theta$ that we use is defined as:
\begin{equation}
\label{eq:LinearNeuron}
    g_\theta(\phi(t),t)_x=\sum_{y,f} K(t)_f W_{d,x,y}\phi(t)_y
\end{equation}
where the time kernel $K(t) \in \mathbb{R}^F$ is given by the first $F$ coefficients of a Fourier expansion, while $W \in \mathbb{R}^{F\times V\times V}$ is a tensor of learnable weights with $V=L\times(R-1)$ being the volume of the active variables of the configurations $\phi(t)$. The divergence of the $g_\theta$ defined in eq.~(\ref{eq:LinearNeuron}) is trivial and is found to be
\begin{equation}
    \bigl(\nabla \cdot g_\theta)(\phi(t),t)=\Tr\biggl[ \sum_{f} K(t)_f W_f \biggr].
\end{equation}
This model corresponds to a continuous in-time extension of the classical linear neuron $y=W\phi$. 

CNFs can be trained so that the learned distribution $q$ well approximates the target $p$  by minimizing the Kullback-Leibler divergence~\cite{kullback1951information}: 
\begin{equation}
\label{eq:dkl}
    D_{KL}(q||p)=\int \DD\phi \; q(\phi) \log \frac{q(\phi)}{p(\phi)}=\int \DD\phi \; q(\phi) \bigl(\log q(\phi)+S_\tmb{NG}(\phi)\bigr)+\log Z
\end{equation}
Since the partition function $Z$ of the target theory is a constant, the problem can be solved by minimizing the first term of eq.~(\ref{eq:dkl}) (also called \textit{variational free-energy}), which provides also an upper bound on $\log Z$.
After training, the partition function of the target $p$ can be computed using the following estimator
\begin{equation}
\label{NF_Z}
    \tilde{Z}=\int \DD\phi \; q(\phi) \tilde{w}(\phi) = \langle \tilde{w}(\phi)\rangle_{\phi \sim q}
\end{equation}
where we introduced the weight
\begin{equation}
\label{flow_weight}
    \tilde{w}=\frac{\exp (-S[\phi])}{q(\phi)}.
\end{equation}
Moreover, an estimator for the expectation value of a generic observable $\mathcal{O}$ can be computed with a reweighting procedure, also referred to as Importance Sampling in the machine learning field~\cite{bishop:2006:PRML, Nicoli:2019gun, Nicoli2021}:
\begin{equation}\label{eq:NF_obs}
    \langle \mathcal{O} \rangle =\frac{1}{Z}\int \DD\phi \; q(\phi) \mathcal{O}(\phi)\Tilde{w}(\phi) = \frac{\langle\mathcal{O}(\phi) \Tilde{w}(\phi)\rangle_{\phi \sim q}}{\langle \Tilde{w}(\phi)\rangle_{\phi \sim q}}.
\end{equation}

\section{Results}
\label{sec:results}

The focus of our numerical study is the computation of both the partition function and the string thickness in two different regimes of the Nambu-Goto theory: high-temperature (HT) ($R \gg L$) and low-temperature (LT) ($L \ll R$). These observables are computed through eq.~(\ref{eq:NF_obs}) using $10^6$ configurations sampled with trained CNFs at fixed $\sigma$, $L$, and $R$. 

For all the models, we use $F=3$ for the temporal kernel of eq.~(\ref{eq:LinearNeuron}) and integrate the NODE over $13$ steps with $T=1$. We initialize the elements of the vector field to $0$ (identity initialization) and train the CNFs for $1000$ Adam~\cite{Kingma:2014vow} iterations with $10000$ samples, $0.0005$ initial learning rate, $\beta_1=0.8$, $\beta_2=0.9$, and a cosine annealing scheduler. 

The performances of the CNFs have been tested using as a metric the so-called Effective Sample Size (ESS)~\cite{Albergo3,Albergo:2021vyo}:
\begin{equation}
 \mbox{ESS} = \frac{\langle \Tilde{w} \rangle^2 }{\langle \Tilde{w}^2 \rangle}
\end{equation}
and serves as a measure of the overlap between the generated probability density $q$ and the target $p$. It takes values between 0 and 1, with the latter representing a perfect training ($q=p$).
We generally use models with ESS~$\geq 0.1$. Due to the limits of the architectures used in this work, we report results for $\sigma > 5.0$ in the HT regime and for $\sigma > 10.0$ in the LT regime. To reduce the computational cost of the simulations, we trained each combination of $L$ and $R$ for $\sigma=5.0,10.0,25.0,100.0$ ($\sigma=5.0$ only for the HT regime); we then used these models as the initialization for CNFs with different string tensions. When performing this transfer learning procedure between models, we train the new CNFs with $100$ Adam iterations with $0.00001$ as the initial learning rate. The error is estimated using a jackknife procedure. 
The code is based on the PyTorch library~\cite{paszke:pytorch} and a simple implementation can be found in~\cite{NFcode}. We ran the computation on Tesla V100 GPUs, with a total cost of approximately $2600$ CPU hours ($1$ GPU hour corresponds to $3$ CPU hours).


\subsection{Simulation setup}

For the high-temperature regime we considered $1804$ combinations of $L$, $R$ and the string tension $\sigma$.
More precisely, we performed simulations for $11$ even values of R between $50$ and $100$, $9$ values of $L$ between $4$ and $12$, $21$ values of $\sigma$ between $5$ and $300$ for $L<8$ and $16$ values between $10$ and $300$ for $L\geq 8$. 

\begin{figure}[ht]
  \centering
\includegraphics[scale=0.8,keepaspectratio=true]{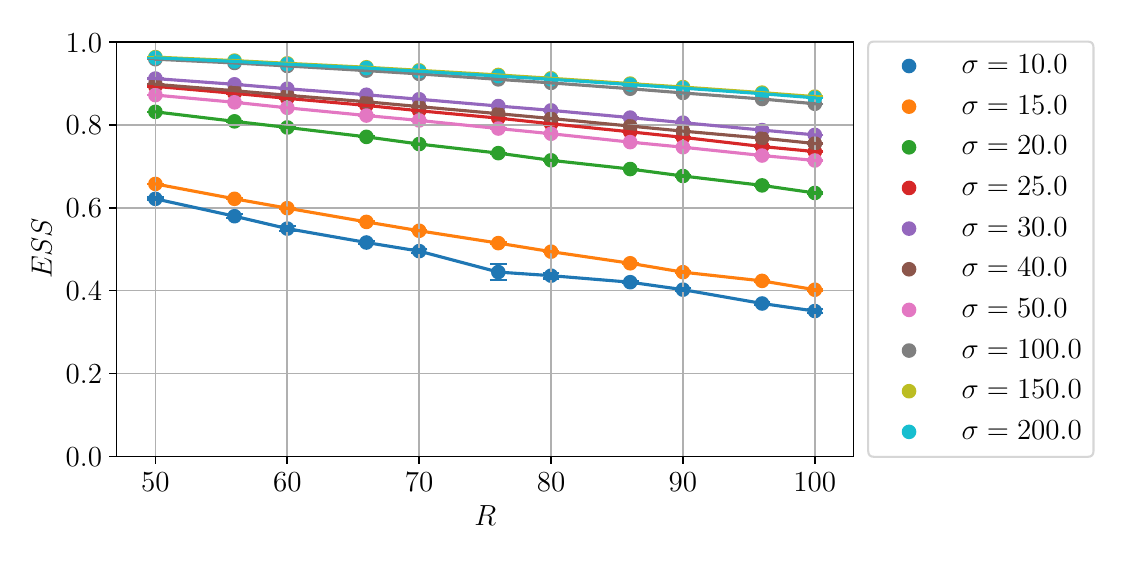}
  \caption{Effective Sample Size as a function of $R$ for fixed $L=10$ and various $\sigma$. Error bars are not visible due to the very small statistical errors.}
   \label{fig:HTESS}
\end{figure}

In fig.~\ref{fig:HTESS}, the ESS is reported as a function of $R$, for different $\sigma$ and $L=10$. This metric indicates that the overlap between the learned and target distribution is very good, even if the scaling with the volume and with $\sigma$ can be a limiting factor for the architectures under consideration in this study. 

\begin{figure}[H]
 \centering
 \includegraphics[scale=0.8,keepaspectratio=true]{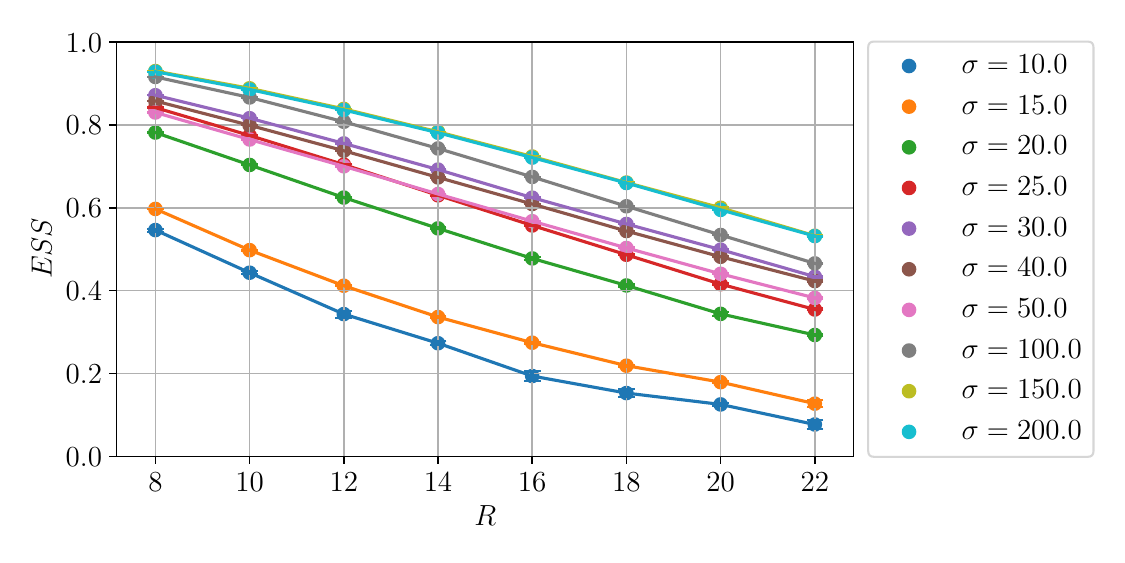}
 \caption{Effective Sample Size as a function of $R$ for $L=90$ and different $\sigma$.}
 \label{fig:LTESS}
\end{figure}

In the low-temperature regime, we performed simulations with $1008$ different combinations of $L$, $R$ and the string tension $\sigma$. We extracted results for eight values of R between $8$ and $22$, six values of $L$ between $84$ and $94$ and $21$ values of $\sigma$ between $10.0$ and $300.0$. 
In fig.~\ref{fig:LTESS} we report the ESS as a function of $R$ for different $\sigma$ and $L=90$. Due to the larger volumes under consideration in this regime the quality of the flows is generally lower than in the HT regime, but it guarantees an efficient sampling of the target distribution nonetheless.

\subsection{Partition Function}
\label{pf}

As mentioned above we benchmark our approach using results for the partition function. Since we focus on the study of the large-$\sigma$ region, we used only data for $\sigma\geq 40$. The most direct check is in the low-temperature regime.

\subsubsection{Low-temperature regime ($L \gg R$)}
  
Following the discussion of section~\ref{sec:NG}, we fitted our data for the partition function for each value of $R$, assuming a linear dependence on $L$ which keeps into account both the area and the perimeter terms; then we looked at the first few terms of the $1/\sigma$ expansion of this linear behaviour
\begin{equation}
 \label{eq:logZLT}
 -\log Z(\sigma,L,R)=\biggl( a^{(0)}_{LT}(R) + \frac{a^{(1)}_{LT}(R)}{\sigma} + \frac{a^{(2)}_{LT}(R)}{\sigma^2} 
 + \frac{a^{(3)}_{LT}(R)}{\sigma^3}\biggr) L ~~.
\end{equation}
Results of these fits are reported in  table~\ref{tab:LTlogZ}.

\begin{table}
\centering
\begin{tabular}{|c c c c c c|} 
 \hline
 $R$ & $a^{(0)}_{LT}(R)$ & $a^{(1)}_{LT}(R)$ & $a^{(2)}_{LT}(R)$ & $a^{(3)}_{LT}(R)$ & $\chi^2/d.o.f.$\\ [0.5ex]
 \hline\hline
8 &  -2.224701(3) & -1.750(1) & 0.5(1)& 2(2) & 1.29  \\
\hline
10 &  -2.893064(5) & -2.280(1) & 0.6(1) & 7(3) & 1.18 \\
\hline
12 &  -3.562534(5) & -2.810(2) & 0.5(1) & 12(4)  & 1.15 \\
\hline
14 &  -4.232608(5) & -3.344(2) & 0.9(1)& 9(4) & 0.74\\
\hline
16 &  -4.903077(6) & -3.878(2) &  1.1(1) & 9(4) & 0.86 \\
\hline
18 &  -5.573818(8) & -4.410(3) & 1.4(2) & 7(6) & 1.28 \\
\hline
20 &  -6.244732(8) & -4.943(3) & 1.7(2) & 4(6) & 1.00 \\
\hline
22 &  -6.91576(1) & -5.480(3) & 2.2(3) & -1(7) & 0.96 \\ 
 \hline
\end{tabular}
\caption{Results for the coefficients of the fit of $\log Z$ of eq.~(\ref{eq:logZLT}) in the LT regime.}
\label{tab:LTlogZ}
\end{table}

Then we used the $a^{(0)}_{LT}(R)$ coefficient to further validate our results, since its behaviour should coincide with that of the lattice discretization of the free bosonic action (i.e. the $\sigma \to \infty$ limit of the Nambu-Goto action). Fitting these values according to
\begin{equation}
\label{eq:logZLT2}
 a^{(0)}_{LT}(R)=A^{(0)}_{LT}R + \frac{B^{(0)}_{LT}}{R}+C^{(0)}_{LT}
\end{equation}
led to values for the constants $A^{(0)}_{LT}$ and  $C^{(0)}_{LT}$ which are compatible with those reported in eq.~(\ref{zfb2}) (see table~\ref{tab:LTlogZ2}).

\begin{figure}[ht]
 \centering
 \includegraphics[scale=0.8,keepaspectratio=true]{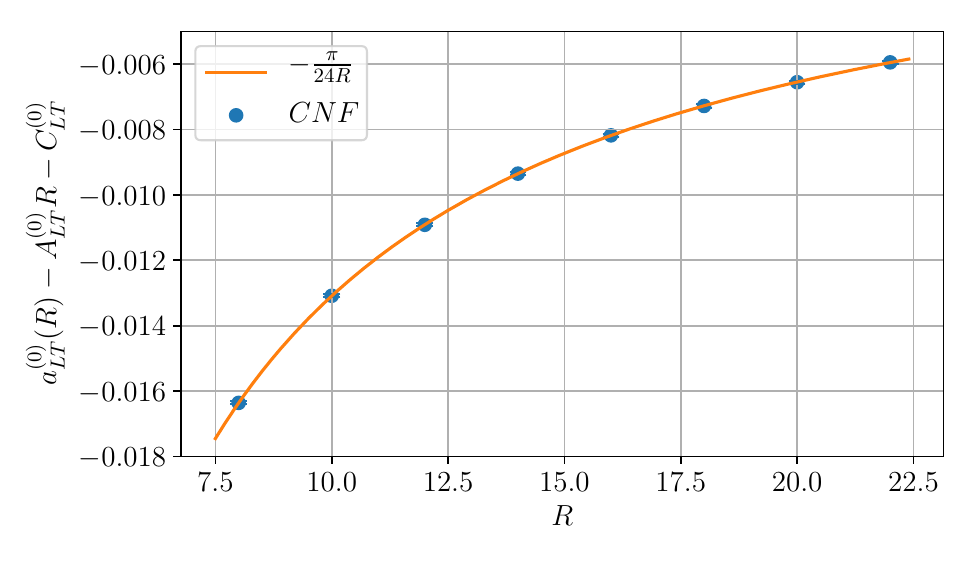}
 \caption{Plot of $a^{(0)}_{LT}(L)-A^{(0)}_{LT}R-C^{(0)}_{LT}$ of eq.~(\ref{eq:logZLT2}) as a function of $R$ compared to the L\"uscher term $-\frac{\pi}{24R}$ (solid line).}
 \label{fig:LTlogZ}
\end{figure}

Moreover, the $B^{(0)}_{LT}$ coefficient is compatible with the coefficient of the L\"uscher term $-\frac{\pi}{24}=-0,13089969...$, which is the only remnant in this limit of the Dedekind function. The agreement with the expectation is clearly visible in fig.~\ref{fig:LTlogZ}, where we plotted the quantity $a^{(0)}_{LT}(R)-A^{(0)}_{LT}R - C^{(0)}_{LT}$ as a function of $R$ and compared it to the L\"uscher term $-\frac{\pi}{24R}$.

A similar analysis can be performed also for the higher order terms of the $1/\sigma$ expansion.  We report for completeness in the second line of tab.\ref{tab:LTlogZ2} the results for $a^{(1)}_{LT}(R)$, fitted with the expression:
 \begin{equation}
\label{eq:logZLT3}
 a^{(1)}_{LT}(R)=A^{(1)}_{LT}R + \frac{B^{(1)}_{LT}}{R^3}+C^{(1)}_{LT}.
\end{equation}

\begin{table}
\centering
\begin{tabular}{|c c c c|} 
 \hline
 $A^{(0)}_{LT}$ & $B^{(0)}_{LT}$ & $C^{(0)}_{LT}$ & $\chi^2/d.o.f.$ \\ [0.5ex]
 \hline\hline
 $-0.335820(2)$ & $-0.1309(2)$ & $  0.47822(4)$ & $0.93$ \\ 
 \hline
\end{tabular}
\\
\vspace{0.3cm}
\begin{tabular}{|c c c c|} 
 \hline
 $A^{(1)}_{LT}$ & $B^{(1)}_{LT}$ & $C^{(1)}_{LT}$ & $\chi^2/d.o.f.$ \\ [0.5ex]
 \hline\hline
 $-0.2669(2)$ & $-5(1)$ & $0.393(4)$ & $0.56$ \\
 \hline
\end{tabular}
\caption{Results for the fit of $a^{(0)}_{LT}(R)$ of eq.~(\ref{eq:logZLT2}) (upper table) and of $a^{(1)}_{LT}(R)$ of eq.~(\ref{eq:logZLT3}) (lower table) in the LT regime.}
\label{tab:LTlogZ2}
\end{table}

\subsubsection{High-temperature regime  ($R \gg L$)}

A non trivial aspect of this regime is that, due to the resummation of the infinite set of exponentials in the Dedekind function in eq.~(\ref{eta}), a logarithmic term appears in the partition function: thus, the analogous of the L\"uscher term has a different coefficient moving from  $-\frac{\pi }{24R}$ to $-\frac{\pi}{6L^2}$ (see eq.~(\ref{eta2bis})).
We first performed a set of preliminary fits keeping the coefficient of the logarithmic term as a free parameter: since the result was compatible with very high precision to $1/2$, we then fixed the coefficient to this value, subtracted the $\log$ term from the data and fitted again with the following function:
\begin{equation}
\label{eq:logZHT}
\begin{split}
-\log Z(\sigma,L,R)-\frac{1}{2}\log \biggl(\frac{2R}{L}\biggr)&   =\biggl(a^{(0)}_{HT}(L)+\frac{a^{(1)}_{HT}(L)}{\sigma}+\frac{a^{(2)}_{HT}(L)}{\sigma^2}+ \frac{a^{(3)}_{HT}(L)}{\sigma^3}\biggr)R +\\ &+ c^{(0)}_{HT}(L) + \frac{c^{(1)}_{HT}(L)}{\sigma} .
\end{split}
\end{equation}
Results are listed in table~\ref{tab:HTlogZ}.

\begin{table}
\centering
\begin{tabular}{|c c c c c c c c|}
 \hline
 L & $a^{(0)}_{HT}(L)$ & $a^{(1)}_{HT}(L)$ & $a^{(2)}_{HT}(L)$  & $a^{(3)}_{HT}(L)$ & $c^{(0)}_{HT}(L)$ & $c^{(1)}_{HT}(L)$ & $\chi^2/d.o.f.$\\ [0.5ex] 
 \hline\hline
4 &  -1.477442(2) & -1.0654(5) & 0.37(4) &  1(1) & 1.9137(1) & 1.61(1) & 0.98 \\
\hline
5 &  -1.785398(3) & -1.3318(6) & 0.52(5) & 0(1) & 2.3921(1) & 1.95(1) & 1.14\\
\hline
6 &  -2.103064(3) & -1.5973(8) & 0.52(6) & 2(1) & 2.8699(2) & 2.34(2) & 1.17 \\
\hline
7 &  -2.426093(4) & -1.8627(9) & 0.58(6) & 3(2) & 3.3488(2) & 2.67(2) & 1.05\\
\hline
8 &  -2.752383(4) & -2.1280(9) & 0.64(7) & 4(2) & 4.3051(2) & 3.43(2) & 1.11 \\
\hline
9 &  -3.080823(4) & -2.3952(9) & 0.77(7) & 4(2) & 4.3051(2) & 3.43(2) & 0.86 \\
\hline
10 &  -3.410762(5) & -2.661(1) & 0.87(9) & 3(2) & 4.7835(2) & 3.79(3) & 1.08 \\
\hline
11 &  -3.741772(5) & -2.928(1) & 1.0(1) & 4(3) & 5.2612(3) & 4.21(3) & 1.29\\
\hline
12 &  -4.073607(5) & -3.194(1) & 1.0(1) & 5(2) & 5.7400(3) & 4.56(3) & 0.99 \\ 
 \hline
\end{tabular}
\caption{Results for the coefficients of the fit of $\log Z$ of eq.~(\ref{eq:logZHT}) in the HT regime.}
\label{tab:HTlogZ}
\end{table}

\begin{figure}[H]
  \centering
\includegraphics[scale=0.9,keepaspectratio=true]{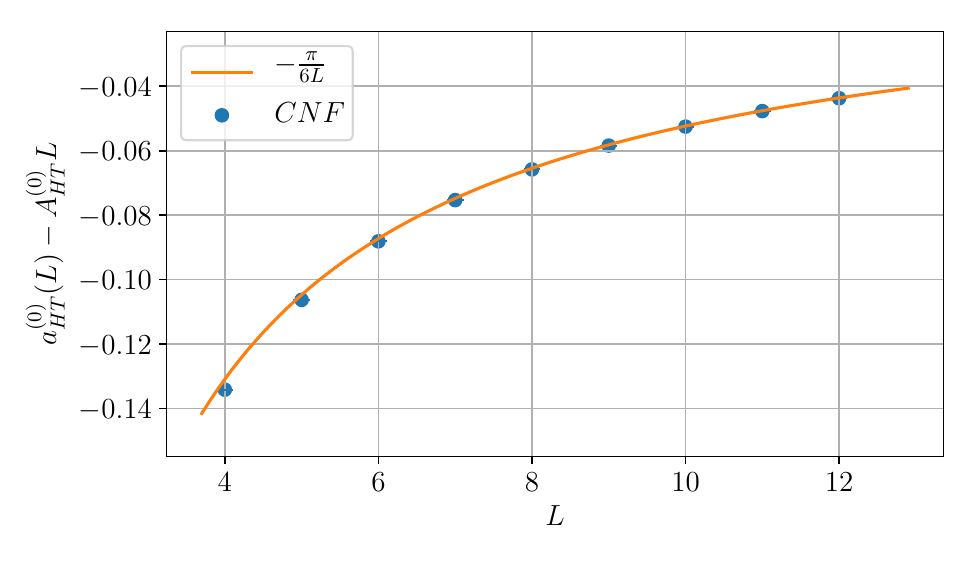}
  \caption{Plot of $a^{(0)}_{HT}(L)-A^{(0)}_{HT}L$ of eq.~(\ref{eq:logZHT2}) as a function of $L$ compared to the Dedekind function prediction $-\frac{\pi}{6L}$ (solid line).}
   \label{fig:HTlogZ}
\end{figure}

Then, in analogy to the low-temperature regime, we fitted the $a^{(0)}_{HT}(L)$ coefficient with
\begin{equation}
\label{eq:logZHT2}
 a^{(0)}_{HT}=A^{(0)}_{HT}L + \frac{B^{(0)}_{HT}}{L} + \frac{B1^{(0)}_{HT}}{L^3} + \frac{B2^{(0)}_{HT}}{L^5}
\end{equation}
and the corresponding results are reported in table~\ref{tab:HTlogZ2}.  
The value of $A^{(0)}_{HT}$ is again compatible within two standard deviations with $A_\tmb{FB}$ and $B^{(0)}_{HT}$ is compatible with the expected value $-\frac{\pi}{6}=-0.523598...$. The quality of this agreement is visible in fig.~\ref{fig:HTlogZ} where $a^{(0)}_{HT}(L)-A^{(0)}_{HT}L$ is plotted as a function of $L$ and compared to the Dedekind function prediction $-\frac{\pi}{6L}$.

\begin{table}
\centering
\begin{tabular}{|c c c c c|} 
 \hline
 $A^{(0)}_{HT}$ & $B^{(0)}_{HT}$ & $B1^{(0)}_{HT}$ & $B2^{(0)}_{HT}$ & $\chi^2/d.o.f.$ \\ [0.5ex]
 \hline\hline
 $-0.335823(2)$ & $-0.5234(2)$ & $-0.179(7)$ & $-0.51(7)$ & $1.89$ \\ 
 \hline
\end{tabular}\\
\vspace{0.3cm}
\begin{tabular}{|c c c|} 
 \hline
 $C^{(0)}_{HT}$ & $D^{(0)}_{HT}$ & $\chi^2/d.o.f.$ \\ [0.5ex] 
 \hline\hline
 $0.47827(3)$ & $0.0007(2)$  & $1.50$ \\ 
 \hline
\end{tabular}\\
\vspace{0.3cm}
\begin{tabular}{|c c c|} 
 \hline
 $A^{(1)}_{HT}$ & $B^{(1)}_{HT}$ & $\chi^2/d.o.f.$ \\ [0.5ex] 
 \hline\hline
 $-0.26612(3)$ & $-0.07(2)$  & $0.43$ \\  
 \hline
\end{tabular}
\caption{Results for the coefficients of the fit of eq.~(\ref{eq:logZHT2}) (upper table), for the coefficients of eq.~(\ref{eq:perHT}) (middle table) and for the coefficients of eq.~(\ref{eq:logZHT3}) (lower table).}
\label{tab:HTlogZ2}
\end{table}

In a similar way we also studied the $L$ dependence of the "perimeter" term  $c^{(0)}_{HT}(L)$ by fitting:
\begin{equation}
    \label{eq:perHT}
    c^{(0)}_{HT}(L)=C^{(0)}_{HT}L+D^{(0)}_{HT}.
\end{equation}
Results are listed in table~\ref{tab:HTlogZ2}, where we see that the dominant term $C^{{(0)}}_{HT}$ is in excellent agreement with the expected value $C_\tmb{FB}$.
We report for completeness also the results of the fit to the $1/\sigma$ term  $a^{(1)}_{HT}(L)$. Using
\begin{equation}
 \label{eq:logZHT3}
 a^{(1)}_{HT}(L)=A^{(1)}_{HT}L + \frac{B^{(1)}_{HT}}{L^3}
\end{equation}
we found the values reported in table~\ref{tab:HTlogZ2}. 


\subsection{Width of the flux tube}

The main goal of this contribution is to show that with a CNF-based simulation it is possible to compute numerically the next-to-leading correction to the flux tube width, which has never been measured before in LGT simulations. At the same time we shall use the agreement of our results with the leading order predictions for the flux tube width as a further sanity check of our approach.
For the analysis of the flux tube width we used all the values of $\sigma$, $R$ and $L$ at our disposal. In the low-temperature regime the next-to-leading corrections are too small and will be below our resolution; however, in the high-temperature regime they are larger and within the precision of our numerical approach. 
 
\subsubsection{Low-temperature regime ($L\gg R$)}

\begin{figure}[H]
  \centering
  \includegraphics[scale=0.8,keepaspectratio=true]{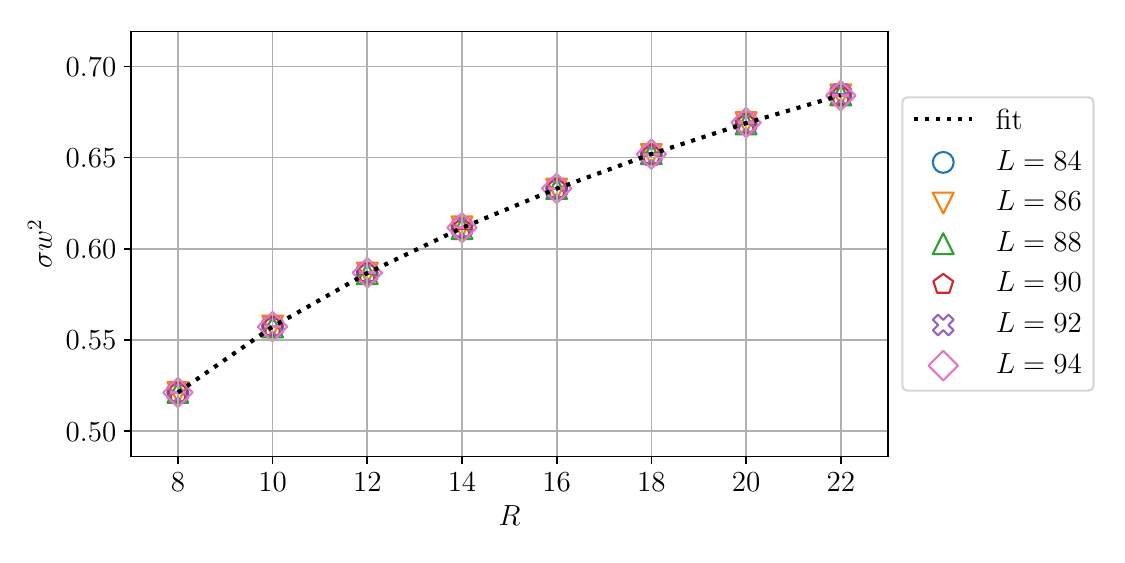}
  \caption{$\sigma w^2$ in the LT regime as a function of $R$ with $\sigma=100.0$ and different $L$. Dotted line represents the best fit of the measures. The error bars are narrower than the points.}
  \label{fig:LTthickness}
\end{figure}

Following the discussion of section~\ref{NGwidth} we fitted our results with
\begin{equation}
\label{eq:w2LT}
 \sigma w^2=\biggl(1+\frac{e^{(0)}_{LT}}{\sigma}+\frac{e^{(1)}_{LT}}{\sigma R^2}\biggr) \biggl(f_{LT}\log(R) + g_{LT}\biggr) + \frac{h^{(0)}_{LT}}{R^2} + \frac{h^{(1)}_{LT}}{\sigma R^2} 
\end{equation}
where we expect (in agreement with eq.~\ref{GPLT})  $f_{LT}=\frac{1}{2\pi}=0.159155...$ for the leading correction, $e^{(1)}_{LT}=-\frac{\pi}{4}=-0.785398...$ and $h^{(1)}_{LT}=\frac{5}{96}=0.05208...$ for the next-to-leading ones. 
The results for the various coefficients are reported in table~\ref{tab:LTthickness}. 
For the leading term we found an excellent agreement with the expected analytical values while, as anticipated, for the next-to-leading ones the statistical uncertainty of the fits is of the same order of the corrections we aim to detect.
In fig.~\ref{fig:LTthickness} we report values of $\sigma w^2$ as a function of $R$ for different $L$ at fixed $\sigma=100.0$, with the corresponding curve from the fit.

\begin{table}
\centering
\begin{tabular}{|c c c c c c c|} 
 \hline
 $e^{(0)}_{LT}$ & $e^{(1)}_{LT}$ & $f_{LT}$ & $g_{LT}$& $h^{0}_{LT}$ & $h^{1}_{LT}$ & $\chi^2/d.o.f.$ \\ [0.5ex] 
 \hline\hline
 1.016(3) & -4(2) & 0.15919(9) & 0.1854(3) & -0.023(6) & 1(1)  & 1.02 \\ 
 \hline
\end{tabular}
\caption{Results for the coefficients of the fit of the string thickness of eq.~(\ref{eq:w2LT}) in the low-temperature regime.}
\label{tab:LTthickness}
\end{table}

\subsubsection{High-temperature regime ($R \gg L$)}
\label{HTwidth}

\begin{figure}[ht]
  \centering
\includegraphics[scale=0.8,keepaspectratio=true]{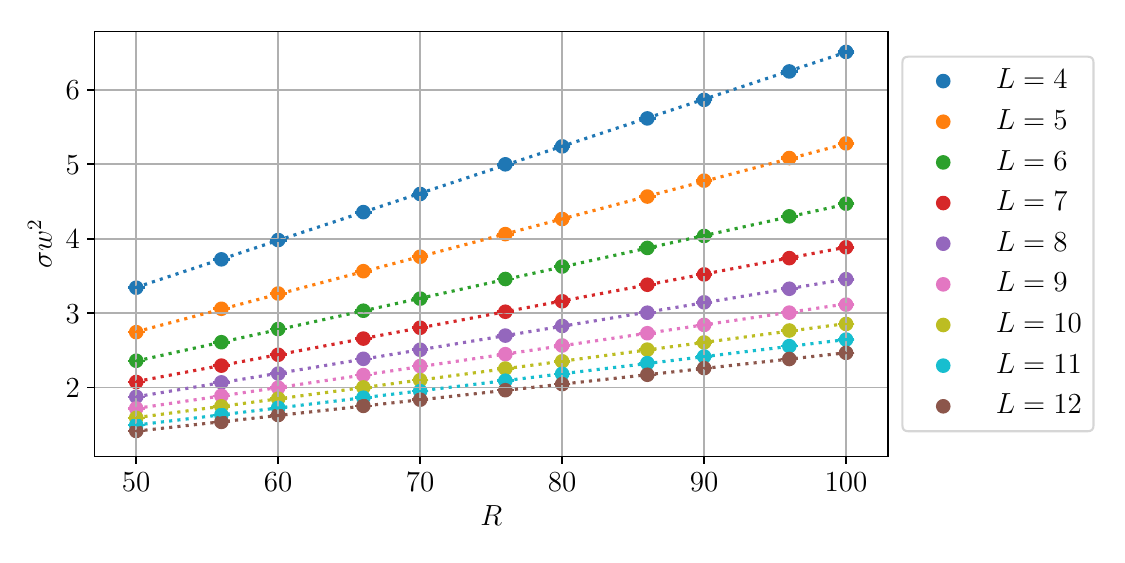}
  \caption{Plot of $\sigma w^2$ in the HT regime as a function of $R$ with $\sigma=100.0$ for several values of $L$. The dotted lines represent the best fit; the error bars are narrower than the points.}
   \label{fig:HTthickness}
 \end{figure}

The situation is more interesting in the high-temperature regime. In agreement with eq.~(\ref{GPHT}) we fitted our data with
\begin{equation}
\label{eq:w2HT}
    \sigma w^2(\sigma,L,R)=\biggl(1+\frac{i^{(0)}_{HT}}{\sigma}+\frac{i^{(1)}_{HT}}{\sigma L^2}\biggr) \biggl(j_{HT}\frac{R}{L} + k_{HT} + l_{HT}\log (L) \biggr)
\end{equation}
where we expect for the leading term $j_{HT}=\frac{1}{4}$, $l_{HT}=\frac{1}{2\pi}=0.159155...$, and for the next-to-leading correction $i^{(1)}_{HT}=\frac{\pi}{6}=0.523598...$; results are reported in table~\ref{tab:HTthickness}. 
We see an excellent agreement with the expected values for all the constants, with a relative error on the next-to-leading correction which is less than 10\%. 
As we mentioned in section~\ref{NGwidth}, a non-trivial behaviour of the flux tube width in the HT regime is its linear increase with the interquark distance $R$. This feature is clearly visible in fig.~\ref{fig:HTthickness}, where we plotted the results of the simulations, together with the best fit curves for a few  selected values of $L$ as a function of $R$. 
In the same figure it is also possible to appreciate the fact that, as $L$ increases, the angular coefficient of the linear term decreases with $1/L$.

\begin{figure}[H]
  \centering
\includegraphics[scale=0.8,keepaspectratio=true]{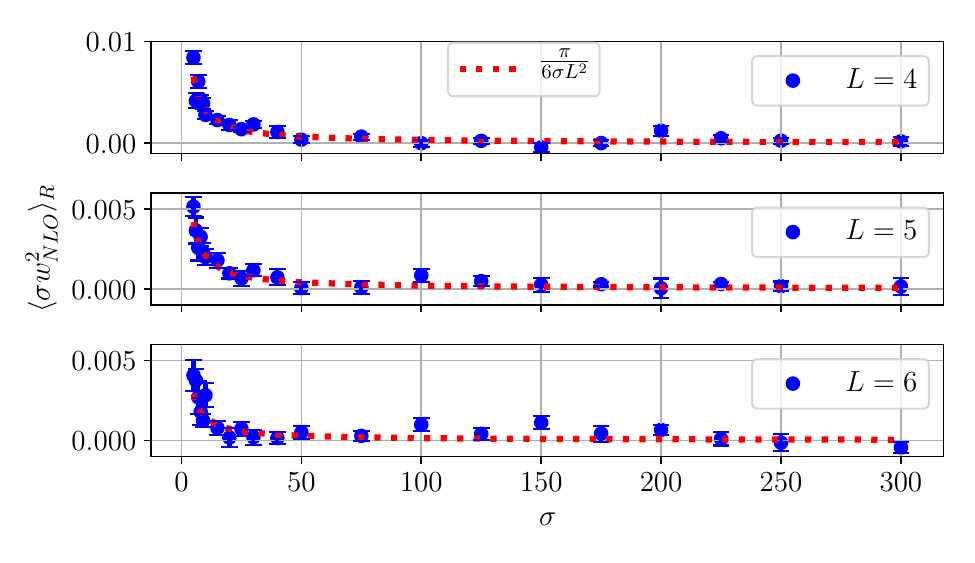}
  \caption{Plot of $\langle \sigma w^2_{NLO}\rangle_R$ as a function of $\sigma$ for three values of $L$ compared to the expected analytical solution $\frac{\pi}{6 \sigma L^2}$.}
   \label{fig:HTPepe}
 \end{figure}

\begin{figure}[H]
  \centering
\includegraphics[scale=0.8,keepaspectratio=true]{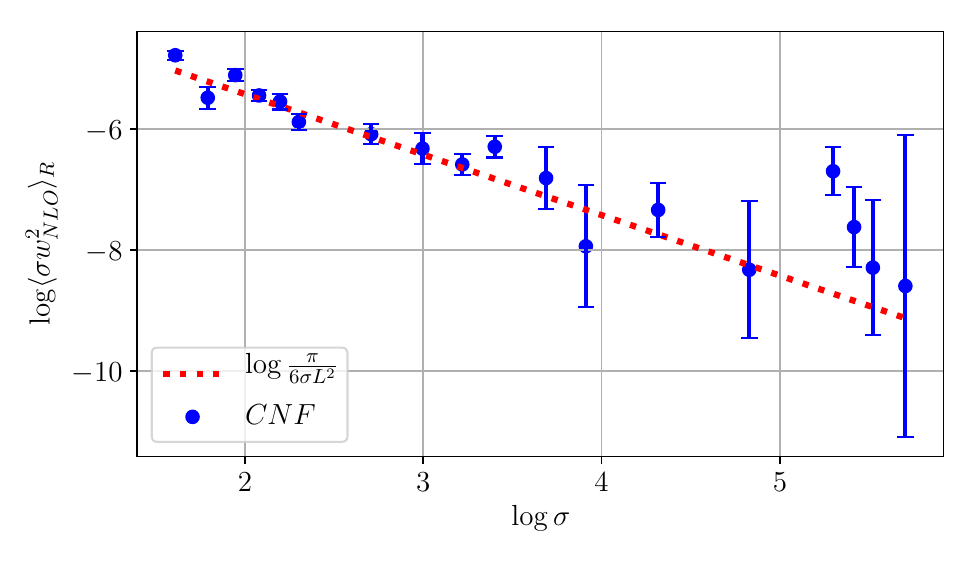}
  \caption{Plot of $\log\langle \sigma w^2_{NLO}\rangle_R$ as a function of $\log\sigma$ for $L=4$ compared to $\log\frac{\pi}{6 \sigma L^2}$.}
   \label{fig:HTPepe2}
 \end{figure}

Then, in order to isolate the next-to-leading contribution, we looked at the following quantity:
\begin{equation}
\langle \sigma w^2_{NLO}\rangle_R(\sigma,L)=\langle\frac{\sigma w^2(\sigma,L,R)}{j_{HT}\frac{R}{L}+k_{HT}+l_{HT}\log (L)}-1-\frac{i^{(0)}_{HT}}{\sigma}\rangle_R,
\end{equation}
where we used the best-fit values for the coefficients $j_{HT}$, $k_{HT}$, $l_{HT}$ and $i^{(0)}_{HT}$ and we took the average over different values of $R$ (as represented by the $\langle ... \rangle_R$ parentheses). Namely, we are assuming to have encoded the whole dependence on $R$ of the width in the $R/L$ term at the denominator. Thus, $\langle \sigma w^2_{NLO}\rangle_R$ should behave as $\frac{\pi}{6 \sigma L^2}$~\cite{Gliozzi:2010jh, Gliozzi:2010zt, Gliozzi:2010zv}: in fig.~\ref{fig:HTPepe} we plot it as a function of $\sigma$ for three values of $L$. The same behaviour for $L=4$ in the $\sigma<50$ region is shown in fig.~\ref{fig:HTPepe2}, together with the expected behaviour of the two-loop calculation~\cite{Gliozzi:2010jh, Gliozzi:2010zt, Gliozzi:2010zv}. The agreement with the theoretical expectation can be easily appreciated in the plots.

\begin{table}
\centering
\begin{tabular}{|c c c c c c|} 
 \hline
 $i^{(0)}_{HT}$ & $i^{(1)}_{HT}$ & $j_{HT}$ & $k_{HT}$ & $l_{HT}$ & $\chi^2/d.o.f.$\\ [0.5ex] 
 \hline\hline
 0.991(2) & 0.55(5) & 0.25007(4) & -0.032(1) & 0.1579(5) & 1.02 \\ 
 \hline
\end{tabular}
\caption{Results for the coefficients of the fit of the string thickness of eq.~(\ref{eq:w2HT}) in the HT regime.}
\label{tab:HTthickness}
\end{table}


\subsection{Comparison with Hybrid Monte~Carlo}
\label{hybrid}
\begin{figure}[H]
  \centering
\includegraphics[scale=0.6,keepaspectratio=true]{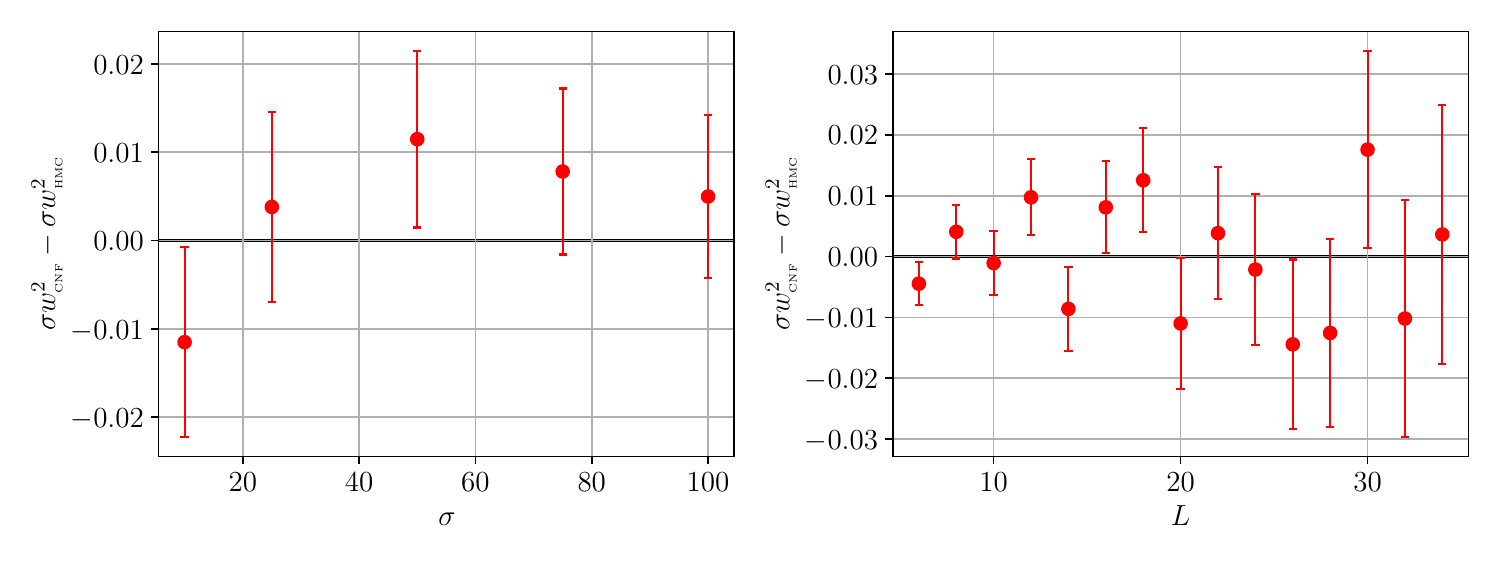}
  \caption{Bias on $\sigma w^2$ between CNFs and HMC from simulations on $L \times R = 20\times 20$ volumes and different $\sigma$ (left plot) and on $L\times L$ lattices with $\sigma=10$ (right plot).}
   \label{fig:BiasHMC}
 \end{figure}
 
In order to further validate the efficiency of our approach, we compared the results obtained with flows with simulations performed using the Hybrid Monte Carlo (HMC) algorithm~\cite{DUANE1987216}. We performed several simulations for various volumes $L \times R$ and $\sigma=10,25,50,75,100$ in order to measure the width $\sigma w^2$. We trained and evaluated the CNFs using the same procedure described at the beginning of this section, whereas for the HMC we generated $100000$ thermalized configurations for each value of $\sigma$. For each molecular dynamics trajectory we fixed $\epsilon=0.1$ and integrated over $10$ steps. We compute errors and autocorrelations for results generated with the HMC algorithm using the $\Gamma-$method~\cite{WOLFF2004143,RAMOS201919} implementation by~\cite{JOSWIG2023108750}. 

\begin{figure}[ht]
  \centering
\includegraphics[scale=0.8,keepaspectratio=true]{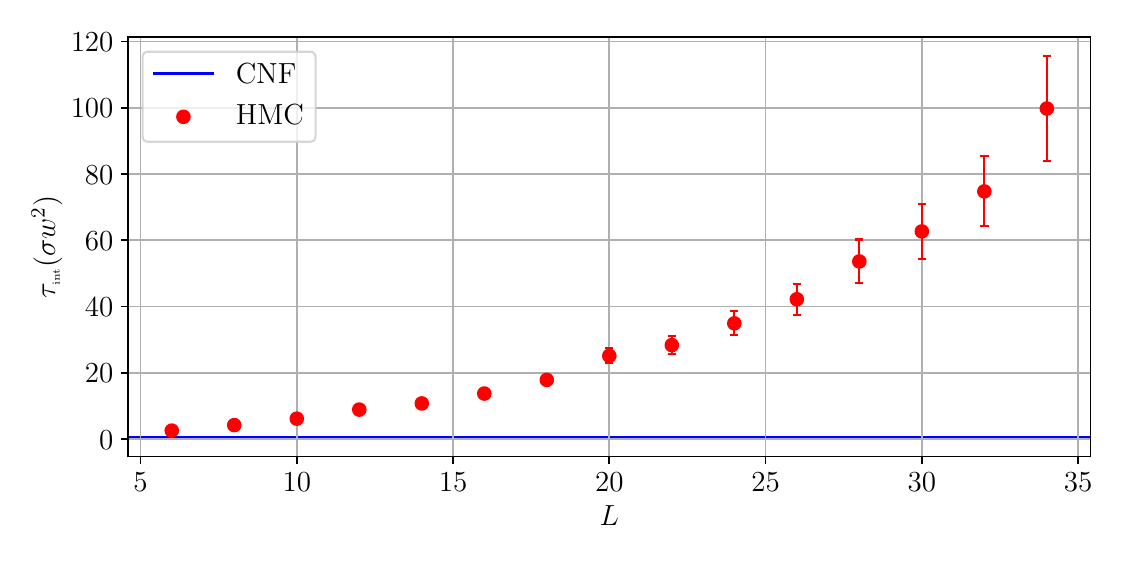}
  \caption{Plot of the integrated auto-correlation time ($\tau_\tmb{int}$) of the HMC algorithm for $L\times L$ lattices with a fixed $\sigma=10$. The solid line denotes the value for uncorrelated configurations.}
   \label{fig:tauint}
 \end{figure}

\begin{figure}[ht]
  \centering
\includegraphics[scale=0.6,keepaspectratio=true]{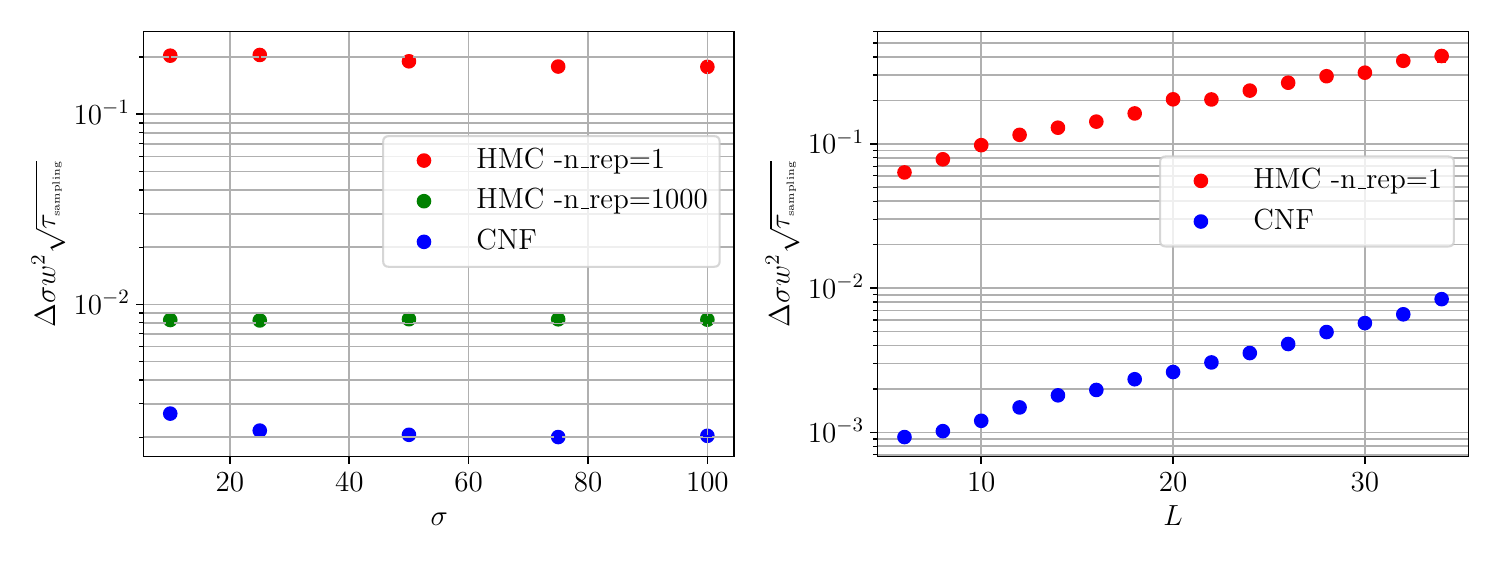}
  \caption{Error on the thickness $\Delta\sigma w^2$ multiplied by the square root of sampling time in a $20 \times 20$ volume and varying $\sigma$ (left plot) and on $L\times L$ lattices with $\sigma=10$ (right plot).}
   \label{fig:ErrorHMC}
 \end{figure}

In fig.~\ref{fig:BiasHMC}, we report the bias between the CNFs and the HMC simulations: the results are fully compatible well within the statistical error both when varying the system size or the string tension. In order to fully appreciate the problematic scaling of the HMC when simulating this model, in fig.~\ref{fig:tauint} we plot the integrated autocorrelation time $\tau_{\mbox{\tiny int}}$ of the width for HMC simulations performed with increasing system sizes. The behaviour is the one typical of a model affected by critical slowing down: when the volume grows, the autocorrelations grow very rapidly. The samples generated with CNFs are independent by construction (and thus $\tau_{\mbox{\tiny int}}=0.5$), but the error on the string width is still affected as the sampling is more difficult. Thus, in fig.~\ref{fig:ErrorHMC} we show the error of both methods multiplied by the square root of the sampling time for a transparent comparison between the cost for the measurements in the two cases. The trained flows appear to be more precise than the HMC simulations by as much as a factor 4 when $1000$ replicas are used to perform HMC simulations, and by about two factors of magnitude when just one replica is used, always keeping the overall statistics fixed. The role of replicas is clearly understood, as they represent independent HMC simulations and are advantageous when large autocorrelations are present in each replica. However, if the number of replicas is increased further thermalization times should be then taken into account in the cost of the sampling. Overall, these results strongly indicate that in the numerical simulation of the Nambu-Goto string CNFs possess a clear advantage with respect to the more traditional Monte~Carlo approach.

Furthermore, we also mention the fact that in traditional Monte~Carlo simulations only ratios of partition functions are directly accessible and their computations via integration methods are notoriously cumbersome. On the other hand, Normalizing Flows offer the possibility (at least in relatively simple models) to compute directly the partition function itself. In light of the results for the flux tube, where the partition function itself has the important physical meaning of the interquark potential, CNFs clearly represent the more efficient option. However, since a careful analysis of the efficiency of the two approaches in computing $\log Z$ is beyond the scope of this work, we refer to~\cite{Nicoli2021} for a rigorous comparison (albeit with a different NF architecture).

\section{Concluding Remarks}

The main goal of this contribution was to show the feasibility of a high precision numerical study of the Nambu-Goto action with a deep generative architecture based on Normalizing Flows. All the terms that we found in the behaviour of the observables under study (the partition function and the width of the flux tube) were already known using the zeta function regularization, but were never observed before in the framework of a lattice regularization of the Nambu-Goto action, due to the intrinsic complexity of performing simulations with traditional methods with such a non-linear action.
As a matter of fact, one of the terms that we studied, i.e. the next-to-leading correction to the flux tube width that we discussed in section~\ref{HTwidth}, had never been observed before even in high-precision lattice gauge theory simulations.

On the numerical side, we sampled thousands of different combinations of $\sigma$, $R$ and $L$. This wide exploration of the phase space of the system was mandatory to reach the needed precision in our analyses. This sampling could be achieved only thanks to the specific properties of NF algorithms: transfer learning, sampling without autocorrelations, easy GPU parallelization. We would like to stress the fact that this class of algorithms proved itself to be much more efficient than the traditional Monte Carlo approach in a setting which is not a simple toy model. 

Furthermore, we see a few natural applications of Normalizing Flow-like methods, that we list below.

\begin{itemize}
\item
They could be used to study the Nambu-Goto action in situations in which the observable of interest is too complex to apply the 
zeta function regularization. This is the case for instance of the higher order corrections of the flux tube width. In this case, in particular, they could be used to test a few existing conjectures on the behaviour of this quantity to all orders in $1/\sigma$ (see for instance~\cite{Caselle:2010zs}).
\item
They could be used to evaluate the contribution of next-to-leading boundary contributions to the effective string action.
\item
They could be used to study corrections to the effective string action beyond Nambu-Goto. In particular a numerical approach of the type discussed in this paper would be mandatory in regimes in which these corrections are large (as we expect to be the case for, say, the $U(1)$ model in three dimensions) and cannot be treated as small perturbations of the Nambu-Goto action.
\item
They could be used to study numerically the $T\bar T$ deformation of the compactified boson, an issue which could be of interest for the description of the spatial string tension in high-temperature LGTs~\cite{Beratto:2019bap}.
\end{itemize}
To conclude, we showed how NFs (and their various generalizations) can already play a crucial role in the understanding of non-perturbative behaviour of Yang-Mills theories through reliable and precise simulations of EST.
In particular, the feasibility of numerical simulations with flows sets the stage for more refined computations in the context of effective string theory actions.

\vskip 1.5cm
\noindent {\large {\bf Acknowledgments}}
\vskip 0.2cm
We thank  M. Panero, S. Bacchio, A. Bulgarelli, K. A. Nicoli and A. Smecca for several insightful discussions. We acknowledge support from the SFT Scientific Initiative of INFN.
This work was partially supported by the Simons Foundation grant
994300 (Simons Collaboration on Confinement and QCD Strings) and by the Prin 2022 grant 2022ZTPK4E. We thank ECT* and the ExtreMe Matter Institute EMMI at GSI, Darmstadt, for support in the framework of an ECT*/EMMI Workshop during which this work has been completed.
\vskip 1cm
\bibliography{biblio}

\end{document}